%% file: main.tex
\documentclass[onecolumn]{IEEEtran}
\usepackage{geometry}
\usepackage{listings}
\usepackage{graphicx}
\usepackage{balance}
\usepackage{color}
\usepackage{url}
\usepackage{nicefrac}
\usepackage{comment}
\usepackage[utf8]{inputenc}
\usepackage{multirow}
\usepackage{booktabs}

\usepackage{amsfonts,amssymb}
\usepackage{bbm,bm}
\usepackage{blkarray}
\usepackage{amsmath}
\usepackage{amsthm}
\usepackage{citesort}
\allowdisplaybreaks

\lstdefinelanguage{JavaScript}{
  keywords={typeof, new, true, false, catch, function, return, null, catch, switch, var, if, in, while, do, else, case, break},
  keywordstyle=\color{blue}\bfseries,
  ndkeywords={class, export, boolean, throw, implements, import, this},
  ndkeywordstyle=\color{darkgray}\bfseries,
  identifierstyle=\color{black},
  sensitive=false,
  comment=[l]{//},
  morecomment=[s]{/*}{*/},
  commentstyle=\color{purple}\ttfamily,
  stringstyle=\color{red}\ttfamily,
  morestring=[b]',
  morestring=[b]"
}

\DeclareMathOperator{\enc}{\encrypt}
\newcommand{\encrypt}{\ensuremath{\mathfrak{E}}}
\newtheorem{defn}{Definition}

\usepackage{algorithm}
\usepackage{algorithmic}

\newtheorem{theorem}{Theorem}
\newtheorem{lemma}{Lemma}

\newtheorem{proposition}{Proposition}

\newcommand{\cupdot}{\mathbin{\mathaccent\cdot\cup}}

\begin{document}

\include{macros}
\allowdisplaybreaks

\title{Collaborative Privacy for Web Applications}

\author{\IEEEauthorblockN{Yihao Hu}
\IEEEauthorblockA{\textit{Electrical and Computer Engineering} \\
\textit{Boston University}\\
yihaoh@bu.edu}
\and
\IEEEauthorblockN{Ari Trachtenberg}
\IEEEauthorblockA{\textit{Electrical and Computer Engineering} \\
\textit{Boston University}\\
trachten@bu.edu}
\and
\IEEEauthorblockN{Prakash Ishwar}
\IEEEauthorblockA{\textit{Electrical and Computer Engineering} \\
\textit{Boston University}\\
pi@bu.edu}
}

\maketitle

 \begin{abstract}
Real-time, online-editing web apps provide free and convenient services for collaboratively editing, sharing and storing files. The benefits of these web applications do not come for free: not only do service providers have full access to the users' files, but they also control access, transmission, and storage mechanisms for them.  As a result, user data may be at risk of data mining, third-party interception, or even manipulation.
To combat this, we propose a new system for helping to preserve the privacy of user data within collaborative environments.  There are several distinct challenges in producing such a system, including developing an encryption mechanism that does not interfere with the back-end (and often proprietary) control mechanisms utilized by the service, and identifying transparent code hooks through which to obfuscate user data.   Toward the first challenge, we develop a character-level encryption scheme that is more resilient to the types of attacks that plague classical substitution ciphers. 
For the second challenge, we design a browser extension that robustly demonstrates
the feasibility of our approach, and show a concrete implementation for Google Chrome and the widely-used Google
Docs platform.  Our example tangibly demonstrates how several users with a shared key can collaboratively and transparently
edit a Google Docs document without revealing the plaintext directly to Google.
\end{abstract}

\section{Introduction}
Collaborative web applications (apps) such as the Google productivity suite (Docs, Sheets, and Slides) enable multiple users to simultaneously edit a number of common documents.  To enable server-side features, such as compression or version control, the contents of these documents are typically available in plaintext to the app provider.  As a result, the provider, affiliated third parties, or malicious parties who have infiltrated the provider, may also be able to mine the plaintext for behavioral advertising, social engineering, or even identity theft.

To help preserve their privacy, some users encrypt their data client-side, allowing only users who know a shared private
key to read the plaintext.  However standard encryption often inhibits the human usability experience~\cite{sheng2006johnny,whitten1999johnny} and its block or streaming encoding is likely to impair or
completely break the collaborative functionality provided by a web service.  Likewise, anonymization overlays like Tor~\cite{tor} or private browsing may only superficially obfuscate the connection between users and data, as the data itself may very well contain deanonymizing features.

In contrast to these existing approaches, we propose a transparent and light-weight encryption layer between clients and providers that protects user data \emph{without} breaking collaborative features.  Users with access to the document secret may view and edit the document within the collaborative framework as if no encryption layer is present.  On the other hand, users who do not know the document secret, and this may include the app provider, see obfuscated text. This layer is implemented through browser extension and it makes extensive use of the standard \texttt{XMLHttpRequest} API~\cite{XMLHttpRequestAPI} used by a variety of web applications (e.g., Google productivity suite, Conceptboard, MeetingWords, Collabedits, Codepen, etc.~\cite{conceptboard,meetingwords,collabedit}) to transmit user edits. 

Our approach is based on a novel character-level variation of the venerable polyalphabetic substitution cipher~\cite{schneier2007applied}.  The benefit of encrypting without the need for context, and at the smallest unit of information of many collaborative apps (\ie one Unicode character), is that our approach maintains functionality and provider bandwidth usage while avoiding heavy-duty reverse-engineering of app-related code or network protocols (which may be obfuscated). 

Though the substitution cipher itself is vulnerable to a number of well-known attacks, such as statistical attacks and chosen plaintext attacks~\cite{sinkov2009elementary}, we provide approaches for strengthening the cipher through standard mitigations approaches such as homophony and mapping randomization in addition to novel approaches based on range extension.  In the latter case, the plaintext is extended from the typically narrow band of the Unicode character space (\eg those associated with the English and/or Greek alphabets) to the entirety of the Unicode space in a manner than helps equalize character and multi-character distributions in order to complicate statistical attacks.  Finally, we demonstrate the effectiveness of our system through the implementation of a Chrome browser extension that showcases its use in preserving privacy for the popular Google Docs collaborative platform.

The following are our \textbf{main contributions}:
\begin{itemize}
\item We identify a robust mechanism for encrypting/decrypting user data within collaborative environments that utilize the \texttt{XMLHttpRequest} API without affecting server-side control traffic.
\item We develop and analyze a novel character-level variation of the polyalphabetic substitution cipher that is more resilient to classical attacks on the cipher.
\item We concretely demonstrate an integration of the two previous contributions as a prototype privacy-preserving Chrome extension for managing Google Docs.
\end{itemize}

We begin in Section~\ref{sec:related} with a review of some of the related work from the literature.  Next we present the architecture of our system in Section~\ref{sec:arch}, including descriptions of our software interface, followed by our new character-level encryption scheme in Section~\ref{subsec:collaborative}.  Section~\ref{sec:demo} describes our prototype together with screenshots of it in action.  We conclude in Section~\ref{sec:conclusions} with some final thoughts, including limitations of our approach.

\section{Related Work}
\label{sec:related}
We next outline several representative (but hardly exhaustive) approaches to web-based privacy preservation from the literature.  Our approach is specifically attuned to online collaborative environments, and our use of a memoryless character-level encryption is thus one key point of departure with the related work below.

\paragraph{M-Aegis}
M-Aegis~\cite{lau2014mimesis} aims to protect data from cloud providers by using a transparent window that sits atop an existing application
and encrypts input data in transit.  Our approach differs from M-Aegis in a number of ways. 
First, M-Aegis focuses on native Android apps,
whereas our approach focuses on browser-hosted apps and is not operating-system dependent.  Rather than  mimicking portions of an app's interface with a GUI overlay, we hook directly into the web application to intercept user input as transparently as possible.
Moreover, in real-time collaborative environments, edits may occur character-by-character and at different locations in a document. Block-based schemes, like those utilized by M-Aegis, require the ability to discern the context of edits in and re-encrypt on the fly.

\paragraph{MessageGuard}
Another closely related work is MessageGuard~\cite{ruoti2017layering}, which implements a system that layers end-to-end encryption on top of existing web applications, using the browser as a global control point and deploying as either a browser extension or a bookmarklet.
MessageGuard uses the \texttt{iFrame} HTML element as a middleware overlay between the user and web app, modifying data before it reaches the application. We, on the other hand, intercept data between the application and server, modifying it in transit.  In this manner, our users' interaction with the app does not change, meaning that their experience is preserved. 

\paragraph{Fully-homomorphic Encryption}
There are also a number of methods of ensuring data confidentiality with the help of the cloud provider, most
notably based on the use of fully-homomorphic encryption (see, for example, ~\cite{gentry2009fully,van2010fully,stehle2010faster,halevi2014helib}), although there is a wealth of additional literature within their citations and reverse-citations).  

These methods aim to
have the server agnostically compute functions of a user's data, and they are aimed toward an honest-but-curious provider.
Our approach does not require any server-side modifications while maintaining transparency to the user or multiple
collaborating users. We also avoid the heavy computational machinery required for these schemes.

\paragraph{Classical Encryption}
More generally, there are quite a number of tools that aim to encrypt user data, as typified by PGP~\cite{zimmermann1995official}
and S/MIME~\cite{ramsdell1999s}. These tools are all meant for one operating user at a time, rather than collaborating
users, and they are not designed up-front to function within the back-end's existing processing methods.

\section{System Architecture}
\label{sec:arch}
Our proposed system has two fundamental components:

\begin{enumerate}
\item A \textbf{browser interface}, which intercepts and modifies data that enters or leaves the app within the browser.
Our specific prototype extension makes use of standard Chrome features to insert interface code between the app and the provider, and, thus, it may be expected to persist over several browser revisions.  Indeed, these features are also common in the popular Firefox browser, and our system should be portable to it as well.
\item a character by character \textbf{collaborative encryption} scheme that runs within the interface to encrypt and decrypt data streams using an extension of the substitution cipher.
\end{enumerate}

We next present details of our architecture, starting in Section~\ref{subsec:threat} with an overview of our threat model.  Section~\ref{sec:browser-interface} describes 
the browser interface together with the software hooks that enable it.  Thereafter, Section~\ref{subsec:collaborative} describes the collaborative encryption scheme, together with analyses and approaches to strengthening its security.

\subsection{Threat Model}
\label{subsec:threat}
We assume that the collaborating users have an out-of-band method for sharing a common secret key for encryption and decryption, and the strength of our encryption scheme is based on some standard assumptions about the statistical properties of the text being edited (elaborated in Section~\ref{subsec:collaborative} within each relevant subsection), which are known to the attacker.

\subsubsection{In Scope}
Our threat model includes an honest-but-curious cloud collaborative service provider or third party that observes and mines data at rest on the service's servers.  Third parties could include attackers with access to the provider's data servers, partners in a business relationship with the provider or law enforcement
agencies.

\subsubsection{Out of Scope}
Since we only focus on the collaborative real-time editors, threats to other kinds of app, such as Facebook Messenger and Emails, are not considered in this paper. Moreover, we primarily protect against attacks at rest on the service provider's servers, and thus do not handle:
\begin{itemize}
\itemit{Browser attacks} - we assume that the browser reliably executes both the application and our browser extension, even though the provider might also provide the browser (e.g. Google and the Google Chrome browser).
\itemit{Side-channel attacks} - either by the provider or by a ``man-in-the-middle'' attacker.  These include active attacks based on statistically correlating key-strokes or client-server communication
with user activities based on fine-grained timing.  We also do not consider information leakage from formatting, style, table structure, or other ``area affects'', and, instead, focus on text alone.  
We believe that these side-channel attacks should be addressed by orthogonal mechanisms.
\itemit{The client-side app} - although we do not assume the app is trustworthy,
we do assume that the implementer of our framework can reverse engineer the application's
protocols to the level of identifying the paths through which input data is transported. Our approach does not cover providers maintaining concealed channels for transferring this data or encrypted metadata, which we would not be expected to access.
\end{itemize}

\subsection{Browser Interface}
\label{sec:browser-interface}
Our approach uses a browser extension-based content script~\cite{mozillaextensions,chromeextensions}
to inject JavaScript payloads into web applications. The payloads hook specific functions
of JavaScript objects that serve as interfaces for app data. With
hooks in place, we can filter and modify the data, which contains
event messages from an app's proprietary protocol.

We have implemented our framework as a browser extension that provides application data interception and modification functionality for the Google Chrome browser. Content scripts typically can access the Document Object Model (DOM)
of targeted pages, but cannot use variables or functions defined by web pages or by other content scripts~\cite{chromeextensions}. 
However, by utilizing features in the environment of the browser, we are able to interact with web
scripts and implement hooks on typical sources and sinks of web application data.

The ability to run code on a web page is only part of the challenge of collaborative encryption.  
Editors that are not HTML-backed editors (like the Google Docs framework).
often generate their client graphical interface through obfuscated JavaScript. As such, a successful prototype must
also identify an appropriate hook through which to intercept communications between the client and the service provider. 
When these frameworks use the \texttt{XMLHttpRequest} Application Program Interface (XHR),
however, one may pick out and overwrite the \texttt{XMLHttpRequest.send} and \texttt{XMLHttpRequest.open}
methods to intercept and modify the entire client-provider data stream.

We next describe some of the details involved with
this prototype implementation, stressing that our content script hooking exploits a
stable feature of the browser (dating back at least to at Chrome 9.0, circa 2011). 

\subsubsection{Content Scripts}
Content scripts run JavaScript code within a specific web page context. 
In Chrome, these scripts may be injected either before the DOM is constructed (\texttt{document\_start} mode), after the DOM is complete (\texttt{document\_end} mode), or right after the window's \texttt{onload} handler is called (\texttt{document\_idle} mode)~\cite{chromeextensions}.  
An enabling feature of these scripts is that, in \texttt{document\_start} mode, they can insert code before
the DOM is constructed.  The result is that, by design, the inserted code overshadows corresponding methods
that are loaded through the web page.

As a general template, the script modeled in Figure~\ref{fig:injection} can be injected into a web page, before the DOM is constructed, to overshadow an existing \texttt{overrideFunction}.  
In our prototype example, we combine two payload injections into a Google Docs page to produce an encryption middleware: 
\begin{itemize}
	\itembf{Outgoing Payload} \\- A script that overwrites \texttt{XMLHttpRequest.send}.
	\itembf{Incoming Payload} \\- A script that overwrites \texttt{XMLHttpRequest.open} and decrypts (with a user-supplied key) initial page content that has been retrieved from the service provider.
\end{itemize}

\begin{figure}
\begin{lstlisting}[language=JavaScript]
  var code = overrideFunction() { 
      ...//Payloads to be injected	
  }; 
  var script = document.createElement("script"); 
  script.textContent = "(" + code + ")();";
  (document.head || 
    document.documentElement).appendChild(script);
\end{lstlisting}
\caption{Injection template.\label{fig:injection}}
\end{figure}

\subsubsection{Outgoing Payload}

\begin{figure}
\begin{lstlisting}[language=JavaScript]
  XMLHttpRequest.prototype.realSend = XMLHttpRequest.prototype.send; 
  var newSend = function(outgoing_data) {
    if (outgoing_data.contain(new_entered_chars)){
      encrypt_algorithm(outgoing_data.new_entered_chars, key);
    }
    this.realSend(outgoing_data);
  };
  XMLHttpRequest.prototype.send = newSend;
\end{lstlisting}
\caption{Outgoing data interception and modification.\label{fig:outgoing}}
\end{figure}

The outgoing interception payload queries the user for an encryption key and then injects a JavaScript snippet similar to that in Figure~\ref{fig:outgoing} into the DOM of the underlying page.
When this snippet redefines the \texttt{XMLHttpRequest.send} method, the new method is subsequently applied to \emph{all} XMLHttpRequest uses and is executed every time \texttt{XMLHttpRequest.send} is called.  In effect, 
the overshadowing method acts as a ``man in the middle'' and allows direct access to the outgoing data so that it may be viewed and modified before being sent out with the original \texttt{XMLHttpRequest.send}.

The \texttt{outgoing\_data} is sent in an incremental fashion, every time changes (such as keystrokes or formatting modifications) are made within the editing window.  In our current prototype, we focus only on modifying keystrokes.

\subsubsection{Incoming Payload}

\begin{figure}
\begin{lstlisting}[language=JavaScript]
    Object.defineProperty(this, "target", {
        get: function(){
            var text = saved_file;
            decrypt(text, key);
            return text;
        },
        set: function(val){
            saved_file = val;
        }});
\end{lstlisting}
\caption{Decoding data stored on the service.\label{fig:incomingRest}}
\end{figure}

The incoming payload requests a decryption key from the user, and then initially decrypts the current state of the document from the service backend using the JavaScript snippet resembling Figure~\ref{fig:incomingRest}. This code is based on the observation that, rather than using \texttt{XMLHttpRequest.open} to access the existing document, Google Docs loads the document content into a page property, and our redefinition
of the getter function of this property allows the Server-stored ciphertext to be intercepted and decrypted into plaintext before being displayed. 

Once the initial state has been established, the incoming payload decrypts updates incoming content from the provider in the fashion of Figure~\ref{fig:incomingUpdates}. Similarly to the overshadowed \texttt{XMLHttpRequest.send} in the outgoing payload, this snippet acts as a ``man in the middle'' to intercept and decrypt incoming data, where \texttt{incoming\_data} here carries only updates to the document. 
The only difference with the overshadowed \texttt{XMLHttp\-Request.send} is that the incoming data is loaded into the property \texttt{responseText}, from which the web app loads updates to the editing window. Therefore, once the getter method is redefined, the incoming data can be successfully intercepted, identified, and modified before being returned to the web app for further processing. 

\begin{figure}
\begin{lstlisting}[language=JavaScript]
  var realOpen = XMLHttpRequest.prototype.open; 
  var newOpen = function(){
    Object.defineProperty(xhr, "responseText", {
      get: function(){
          if (xhr.readyState===4){
            var incoming_data = xhr.response;
            if (incoming_data.contains(new_inputs))
            decrypt(incoming_data.new_inputs);
            return incoming_data;
          } }
    });
    realOpen.apply(this, arguments);
  };
  XMLHttpRequest.prototype.open = newOpen;
\end{lstlisting}
\caption{Decoding incoming user updates.\label{fig:incomingUpdates}}
\end{figure}

\section{Collaborative Encryption}
\label{subsec:collaborative}
Traditional encryption schemes aim to ``confuse and diffuse'' a plaintext~\cite{shannon1949communication}
into a ciphertext, so that a small perturbation in the plaintext produces an unpredictable
``avalanche''~\cite{feistel1973cryptography} of changes in the ciphertext.  As an overlay
for a collaborative system, however, this model has some significant drawbacks.

Consider, for example, several users editing a shared document online.  If one user changes an ``e'' to an ``a'' somewhere
in the document, the cloud back-end propagates only this change to the other users, and not an entirely new copy of the document, in order to limit communication overhead.  From the perspective of an overlay, however,
if the one letter change completely affects an encryption block, then, in effect, 
the back-end must update all users with the entire block that was changed upon every edit, and the collaboration is
very inefficient.

As such, for our platform we seek a ``locally-encodable'' encryption scheme that manages two seemingly contradictory demands:
\begin{enumerate}
\item Minimize the number of ciphertext characters affected by a small change to the plaintext.
\item Make it difficult to determine a plaintext, or even parts of a plaintext, from a given ciphertext.
\end{enumerate}

The second demand is typical of encryption protocols, and can be defined in a number of ways, most notably based on computational or information-theoretic assumptions.  The first demand is specific to our collaborative context, and we next develop several approaches for meeting it.

The overarching basis for our approach will be the classical substitution cipher, as formalized and described in
Section~\ref{sec:substitution}.  It is well known that the substitution cipher leaks statistical information about its plaintext and is also not robust to a (chosen or known) plaintext attack.  For the issue of statistical leakage, we propose two approaches based on spreading the plaintext alphabet
over a larger ciphertext alphabet.  In Section~\ref{sec:extension}, we consider the approach of apportioning the plaintext
alphabet into many equal-sized blocks in the ciphertext alphabet, a practical scheme with a challenging analysis.  In
Section~\ref{sec:infotheory}, on the other hand, we evaluate apportioning the plaintext into varying-sized blocks in the
ciphertext alphabet, resulting in a more complicated implementation with a simpler analysis.  Finally, in Section~\ref{subsec:plaintext} we consider mitigations for plaintext attacks.

\subsection{Substitution - a simple approach}
\label{sec:substitution}
We formalize the first demand of our locally-encodable encryption as
Definition~\ref{def:smallDelta}, based on an encryption function $\enc_{k}:\Sigma^* \longrightarrow \Sigma^*$
indexed by a key string $k \in \Sigma^*$ (which is the encryption secret shared by collaborating users) and mapping plaintext strings
over an alphabet $\Sigma$ into ciphertext strings over the same alphabet.

\begin{defn}
\label{def:smallDelta}
An encryption function $\enc$ is \emph{locally-encodable} if, for some constants $c \geq 1$ and all $k,s,s' \in \Sigma^*$,
\[
\delta \left( \enc_k(s),\enc_k(s') \right) \leq c\, \delta (s, s'), 
\]
where $\delta(x,y)$ is the Levenshtein edit distance metric~\cite{levenshtein1966binary},
denoting the minimum number of insertions, deletions,
and/or single character transpositions needed to transform string $x$ into string $y$.
\end{defn}

When $c$ is the ciphertext length, Definition~\ref{def:smallDelta} generalizes any deterministic, fixed-length
encryption algorithm. Likewise, $c<1$ is disallowed because it prohibits unique decryption, in that two different plaintexts might map
to the same ciphertext.

From the perspective of interfacing cleanly with existing collaborative environments, we desire a position  encryption is one that is position independent.

\begin{defn}
\label{def:posInd}
Encryption $\enc$ is \emph{position-independent} if:
\[
\enc_k(s) = \enc_k(s_0)+\enc_k(s_1)+\enc_k(s_2)+\ldots+\enc_k(s_n),
\]
where $+$ denotes concatenation and $s=s_0+s_1+\ldots+s_n$.
\end{defn}

Position-independent encryptions are useful because they can be calculated
without needing to consider the entire text.  More precisely, the encryption
can be calculated based on individual characters being edited.

Consider, for example, a plaintext $s=\mbox{\texttt{p\textcolor{blue}{h}ilosophi\textcolor{blue}{c}al}}$ in a document that is being edited
collaboratively, and suppose the string is encrypted with a position-independent
scheme as $\enc_k(s)=\mbox{\texttt{K\textcolor{blue}{S}ROLHLKSR\textcolor{blue}{X}ZO}}$ on the server.  Changing the plaintext
by transposing the ``\texttt{c}'' to a ``\texttt{g}'' and deleting first ``\texttt{h}'' corresponds to
modifying the ciphertext by transposing $\enc_k(\mbox{c})=\mbox{\texttt{X}}$ to $\enc_k(\mbox{g})=\mbox{\texttt{T}}$
and deleting $\enc_k(\mbox{\texttt{h}})=\mbox{\texttt{S}}$.  The implementational consequence of this is that,
when looking at the event messages that are being sent between users, we just need to identify
the actual letters being transmitted, and not other metadata, such as their position in the text.

It is not hard to see that the simple \emph{substitution cipher}, which maps strings based on
a one-to-one correspondence between input and output character spaces, is an encryption scheme
that is both locally-encodable ($c=1$) and position-independent.  One of its well-known
drawbacks is that, despite the theoretically large work-factor to break it (95! for printable ASCII
characters), the cipher readily yields to classical statistical analysis, since it preserves
the character distribution of its source and leaks exact information about where a given string is being changed.

\begin{figure}
\centering
\includegraphics[width=\columnwidth]{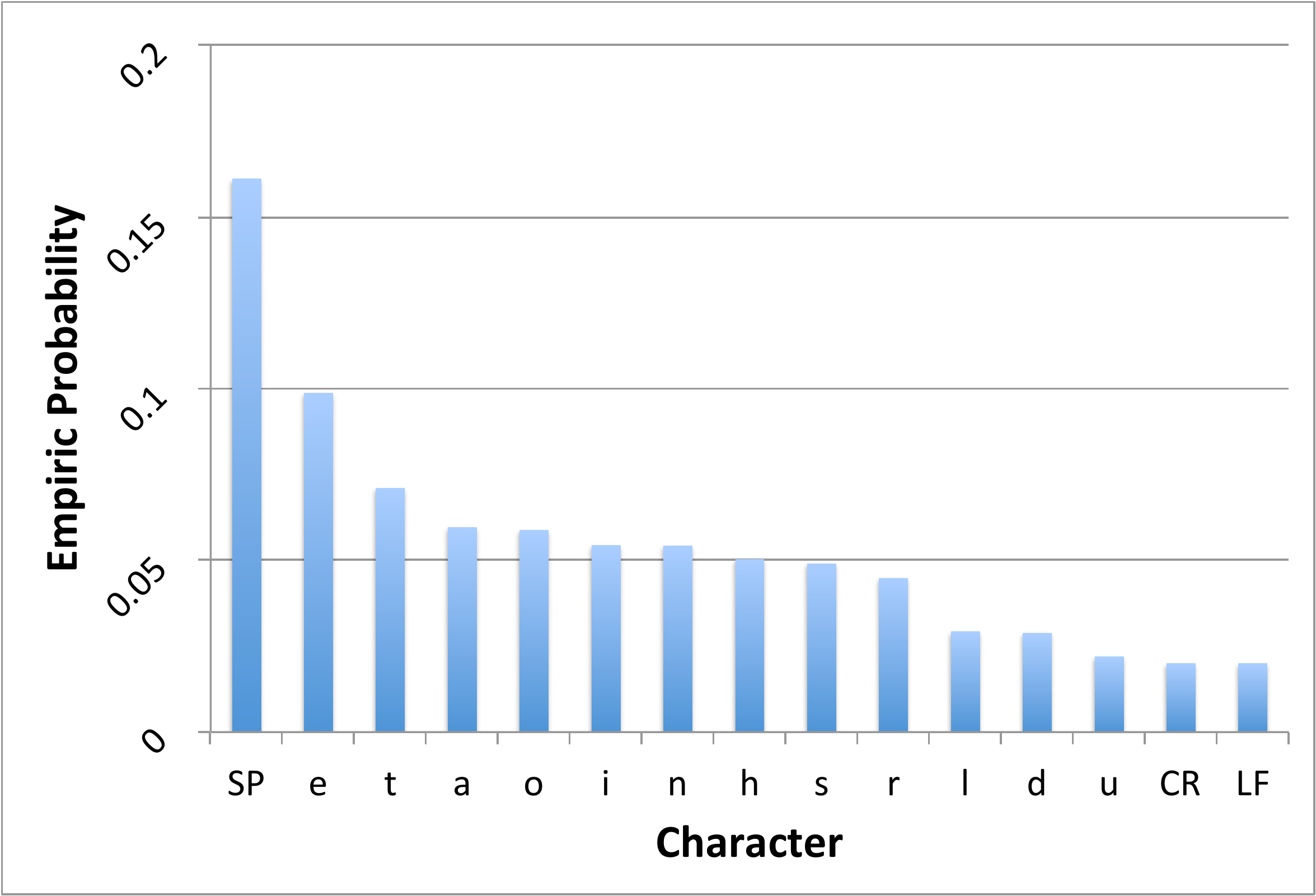}
\caption{Histogram of characters in Gutenberg's translation of Mark Twain's ``Tom Sawyer''.  \texttt{SP}, \texttt{CR}, and
\texttt{LF} denote a space, carriage-return, and linefeed, respectively.\label{fig:plaintext}}
\vglue -3ex
\end{figure}

Figure~\ref{fig:plaintext} shows a histogram of the characters found in Mark Twain's \emph{The Adventures
of Tom Sawyer}~\cite{TomSawyer} as a baseline for subsequent examples.  One can clearly see the
dominance of characters such as the space (\texttt{SP}) and letter \texttt{e}, which can thus
be identified in the substitution-encrypted text.

\subsection{Alphabet Extension}
\label{sec:extension}
To combat frequency analysis, it is possible to embed the range of usable characters (say, the 95 printable ASCII characters) within the larger Unicode space that is supported by many web applications.  For JavaScript engines, it is convenient to use the Unicode characters in the range \texttt{0x0020} to \texttt{0xD7FF}, since many JavaScript engines encode strings as sequences of 16-bit Unicode Transformation Format (UTF-16) code units, where each character is represented by a single code unit.

\subsubsection{Encryption}
One way of achieving this embedding is to divide the available Unicode region into non-intersecting $95$-character blocks
(corresponding to printable ASCII characters), and assigning to each block a pseudorandom
permutation seeded by the block's ID and the encryption key (\ie the
password shared by the various users).  To encrypt a printable character, one uniformly
randomly picks a $95$-character block from the Unicode range, and uses the corresponding
permutation to produce a Unicode character.

As an example, consider the extension algorithm encoding a plaintext character \texttt{b}.  The user picks a $95$-character Unicode block in a random (and not necessarily reproducible) way;  in this case, she may choose the second block, with Unicode characters in the range
\texttt{0x007F}-\texttt{0x00DD}.  A concatenation of the block ID (2) and a shared password is then used to seed
a pseudorandom number generator (PRNG) that produces a permutation of the Unicode characters in the
range; there are a number of well-known and efficient methods for producing such a random permutation of
$k$ integers, dating back (at least) to Hall and
Knuth (see ~\cite{hall1965combinatorial,lehmer1960teaching,sedgewick1977permutation} for some implementations).
Since our plaintext \texttt{b} is the 66th printable ASCII character, we replace it with the 66th element of
our pseudorandom permutation, in this case \texttt{{<}{<}}, which is our ciphertext character.

\subsubsection{Decryption}
To decrypt a ciphertext, a second (authorized) user identifies the Unicode block in which
the encrypted character is found, and seeds a PRNG with a concatenation of the resulting block ID and the
shared password.  This PRNG is then used to produce a pseudorandom permutation, the same one produced
by the encrypting user, which is inverted to produce the original plaintext character.

In our earlier example, the second user would identify ciphertext \texttt{{<}{<}} as belonging
to Unicode block \texttt{0x007F}-\texttt{0x00DD}, which has block ID (2).  She would concatenate this ID with
her shared password to seed a PRNG and produce the permutation found in that block on the figure.  The
permutation is a one-to-one correspondence between printable ASCII characters and Unicode characters
in the range, so it is readily inverted to produce the plaintext \texttt{b}.

\subsubsection{Unigram analysis}
This approach naturally increases the entropy of the resulting ciphertext over simple substitution, as expressed by the following straightforward lemma (we use the notation \M\ to denote the plaintext alphabet, and \C\ to denote the ciphertext alphabet).

\begin{lemma}
Extending the base alphabet from $|\M|$ 
to $|\C|$ characters in this manner increases character entropy by $\log_2 \left( \frac{|\C|}{|\M|} \right)$ bits.
\end{lemma}
The plaintext unigram entropy is given by $H(P) = - \sum_{i \in \M} p_i \log_2 (p_i).$
The encoding process uniformly distributes characters among the $\gamma = \frac{|\C|}{|\M|}$
blocks, meaning that the probability of seeing a character $\epsilon$ corresponding
to $i$ (according to the random permutation of its block) in the ciphertext alphabet
is $\Pr(\epsilon) = \frac{p_i}{\gamma}$.  Computing the resulting entropy of the
ciphertext $C$ produces:
\begin{align*}
H(C) &= -\sum_{\epsilon \in \C} \Pr(\epsilon) \log_2 \left( \Pr(\epsilon) \right)
= - \gamma \sum_{i \in \M} \frac{p_i}{\gamma} \log_2 \left( \frac{p_i}{\gamma} \right) \\
&= H(P) + \log_2 (\gamma). \qquad \qquad \qquad \qquad \qquad \qquad \qquad 
\end{align*}

In the specific case of graphic Unicode characters consistently accessible
via JavaScript
(i.e., \texttt{0x0020} to \texttt{0xD7FF}),  we add roughly $9$ bits of entropy to the ciphertext.
Extending to $1,112,064$
valid code points of UTF-8 provides approximately $13$ bits of extra entropy.

In the sorted empirical entropies of the blocks
produced by this extended encoding of ``Tom Sawyer''.  For example, all blocks
have a reasonably high entropy within roughly $0.2$
bits of the input text's entropy of $4.61$ bits,
and the overall ciphertext entropy is roughly $13.73$, 
which is about $9$ bits more than the input text entropy.   

\begin{figure}[t]
\centering
\includegraphics[width=\columnwidth]{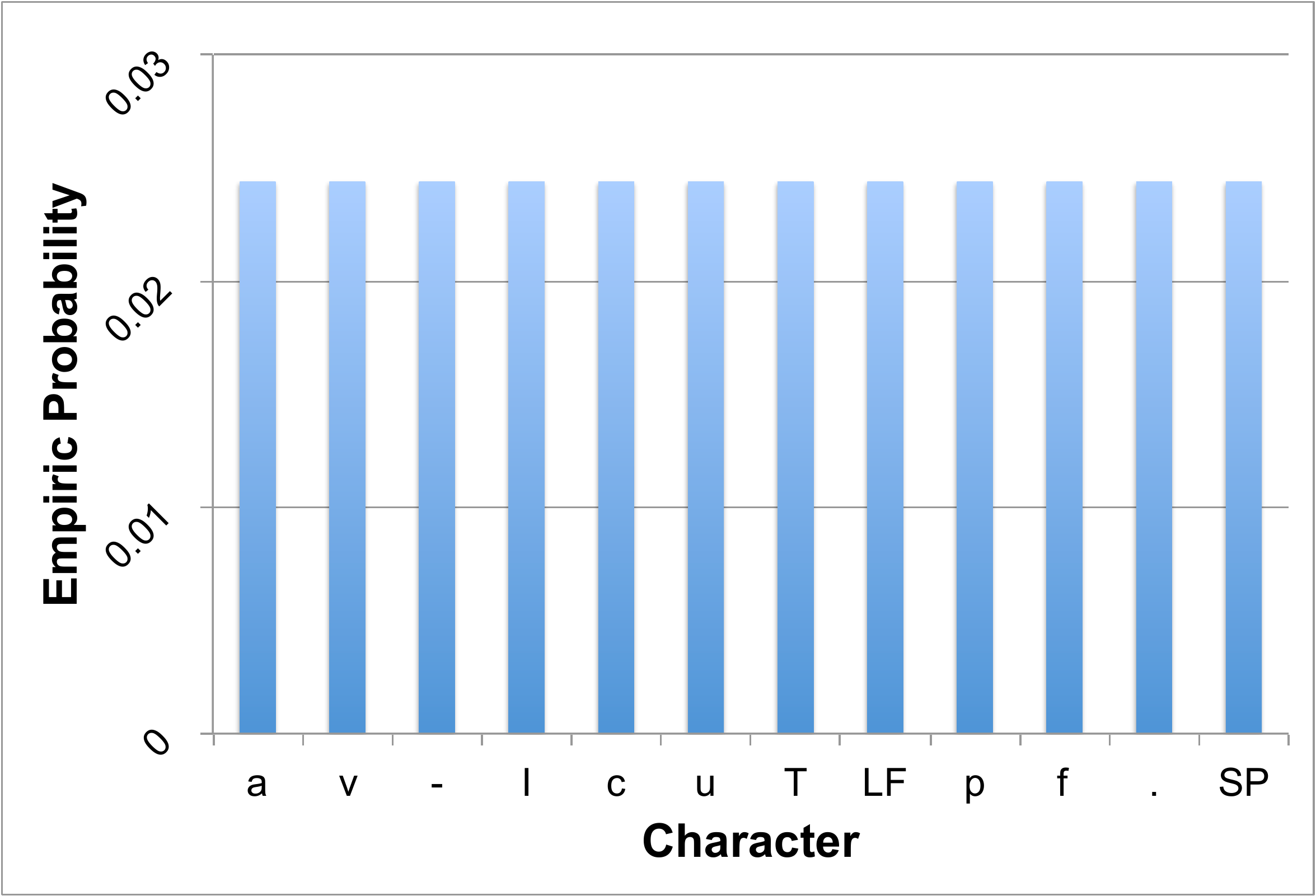}
\caption{Histogram of plaintext characters mapping to Unicode characters \texttt{0x18DF0} - \texttt{0x18E4F} in the \emph{entropy maximization} cipher.\label{fig:entropyFull_block}}
\vglue -2ex
\end{figure}

The problem with this scheme becomes evident by examining the histogram of
input characters mapped to a specific block,
wherein uneven distribution of plaintext characters may carry over to   
ciphertext characters; for example, one may readily identify that
the most common ciphertext characters will be mapped from a space character and
the letter ``\texttt{e}''.

\subsubsection{Greedy Entropy Maximization}
Within the alphabet extension scheme of Section~\ref{sec:extension} there is flexibility
about which Unicode block range to use in producing a ciphertext character.  Though a random choice
produces a high overall entropy, it might be more advantageous to flatten the histogram
of characters mapped into each block. 

A joint optimization of entropy across the entire plaintext message is inappropriate for our application, which requires online
encryption one character at a time, but a greedy entropy maximization is
feasible.  In this approach, we maintain the histograms of each Unicode block in memory.  When
a new character needs to be mapped, we consider its effect on the entropy
of each Unicode block and add it to the block for which it most raises the entropy, breaking
ties uniformly at random.  This has the effect of significantly flattening the distribution of characters.

Indeed, Figure~\ref{fig:entropyFull_block} shows that the first $200$k of characters from the same text under
heuristic produces a flat distribution for a given block 
, with
correspondingly high entropies for each block. 

\paragraph{Local decodability for $c>1$}
By increasing the constant $c$ in the locally-decodable definition, it is possible
to even further reduce the amount of information leaked by an edit, at the expense
of significantly increasing the implementational complexity of the system.  In
this scenario, one user edit results in a constant number of edits in the ciphertext.
A consequence of this approach is that the overlay has to be able to produce edit
events \emph{de novo}, which is fragile to updates in the underlying web application.

\paragraph{Security}
Our encryption scheme relies on two elements for its security.  First, by greedily
maximizing entropy, we end up significantly flattening the distribution of characters
mapped into a given block, complicating character-based frequency analysis.  Second,
the pseudo-random permutation choices per block provide 
some separability, in that decoding the permutation of one block does not directly
lead to the decoding of another block (although it may provide side-information with
which to mount an attack).

\subsection{Information Theoretic Optimization}
\label{sec:infotheory}
Thus far, we have utilized fixed-length blocks, each encrypting (through
substitution) the entire range of usable plaintext characters and a heuristic greedy entropy maximization method to flatten the block-wise unigram distribution.  However, variable-length blocks, with lengths adapted to the probability distribution of plaintext characters, can provide even better defense against statistical attacks because the unigram distribution of ciphertext characters can be made arbitrarily close to uniform. \\

\subsubsection{Variable-length block algorithm} Let $\mathcal{M}$ denote the plaintext alphabet of size $m$. Each
plaintext character $i \in \mathcal{M}$ is first mapped (independently of other plaintext characters) uniformly at
random to a ciphertext character $X_i \in \mathcal{C}_i$, where
$\mathcal{C}_1,\ldots,\mathcal{C}_m$ are $m$ disjoint ciphertext
sub-alphabets (one for each plaintext character) and $\mathcal{C} :=
{\cupdot}_{i=1}^m \mathcal{C}_i$ is the entire ciphertext alphabet.
Let $\sigma_{\mathcal{C}}$ denote a permutation on $\mathcal{C}$. Our
randomized homophonic substitution cipher can be described by the encryption function which $i \longrightarrow \sigma_{\mathcal{C}}(X_i)$, which is invertible with knowledge of $\sigma_{\mathcal{C}}$.
Here, the permutation represents a shared secret key which is
available to both the encryption and decryption algorithms, but not
the attacker.

\par In order to simplify the exposition, we shall assume that the plaintext stochastic process 
$Z_1,\ldots,Z_n$ is first-order stationary, meaning that the unigram (i.e., marginal) distribution of individual plaintext characters is the same at all positions within the plaintext sequence. 
While this assumption may not hold exactly in practice, it is a fairly weak technical assumption to make since it still allows the process to be non-stationarity (of higher orders) and also have strong temporal dependencies (memory). Moreover, it can be made to hold to any desired degree by encrypting a suitably long sequence of consecutive plaintext characters at once as a group and permuting the {\it sequential ordering} of characters within the group using another shared secret key. For simplicity, however, we will assume that the first-order stationarity condition holds without such grouping and sequential ordering permutation. \\

\subsubsection{Unigram distribution} Since the encryption process operates in a character-wise and statistically time-invariant manner, the ciphertext character process is also first-order stationary. If the unigram (first-order) probability mass function (pmf) of the plaintext is $\Pr(Z_1 = i) = p_i$, $i \in \mathcal{M}$, then the unigram (first-order) pmf of the ciphertext is given by $q_j := \Pr(\sigma_{\mathcal{C}}(X_{Z_1}) = j) = \frac{p_i}{|\mathcal{C}_i|}$, for all $i\in \mathcal{M}$ and $j\in \sigma_{\mathcal{C}}(\mathcal{C}_i)$. 
\[
\forall i\in \mathcal{M} \mbox{ and all } j\in \sigma_{\mathcal{C}}(\mathcal{C}_i),
q_j := \Pr(\sigma_{\mathcal{C}}(X_{Z_1}) = j) = \frac{p_i}{|\mathcal{C}_i|}.
\]
This is because in order to get ciphertext character $j$, the plaintext character $i$ that corresponds to it must be generated (this happens with probability $p_i$) and then the particular ciphertext character within the bin $\mathcal{C}_i$ from which $j$ arises (under permutation $\sigma_{\mathcal{C}}$) must be picked (this happens with probability $\nicefrac{1}{|\mathcal{C}_i|}$).

\begin{proposition} If the unigram pmf over plaintext characters and the ciphertext
  sub-alphabet sizes are such that for all $i \in \mathcal{M}$, $|\mathcal{C}_i| = p_i\cdot |\mathcal{C}|$,
then the unigram distribution of ciphertext characters is exactly uniform over the ciphertext alphabet, i.e., $q_j = \nicefrac{1}{|\mathcal{C}|}$ for all $j\in \mathcal{C}$.
\end{proposition}

If the plaintext unigram probabilities $p_i$ are all rational numbers, then the ciphertext unigram probabilities $q_j$ can be made {\it exactly} uniform over the ciphertext alphabet $\mathcal{C}$ using a sufficiently large, but finite, $|\mathcal{C}|$. In practice, the plaintext unigram probabilities would be estimated empirically as normalized character counts (frequencies) in some corpus of documents. The estimated probabilities would therefore be rational numbers.
If, on the other hand, even one $p_i$ is irrational, exact uniformity cannot be attained with
any finite $|\mathcal{C}|$. However, one can always approximate any irrational
fraction with a rational one with a sufficiently large denominator. 
Thus, the ciphertext unigram distribution can be made as close to uniform as desired by making
$|\mathcal{C}|$ sufficiently large. In practice, if $p_i\cdot |\mathcal{C}|, i \in \mathcal{M}$ are not integers, we would drop the fractional parts and distribute (in some manner) any remaining characters in the ciphertext alphabet among the plaintext alphabet characters.

In different
scenarios, the plaintext distribution may be known to both the users
and the attacker, or only to the various users, or to none. Similarly,
the {\it sizes} of the ciphertext sub-alphabets
$|\mathcal{C}_1|,\ldots,|\mathcal{C}_m|$ may be known to both the
users and the attacker or only the users. However, the overall
ciphertext alphabet $\mathcal{C}$ will be known to both the users and
the attacker. In what follows, we assume that the users know everything 
and with the exception of the secret key, the attacker also knows everything.

\par If the unigram ciphertext distribution $q_j$ can be made exactly uniform,
then no statistical test based only on observed ciphertext 
(ciphertext-only attack) will be able to tease the plaintext characters apart (with confidence better than a random guess) using 
only a unigram frequency analysis.
On the other hand if the $q_j$'s are not exactly uniform, and they are all distinct 
for different $j$ and known to the attacker, then as $n$ becomes very large, a ciphertext-only attack may be able break the cipher with overwhelming probability. 
However, the closer that the $q_j$'s are to being 
uniform, the longer that the attacker will have to wait to gather enough ciphertext characters before being able to break the cipher with sufficient confidence. \\

\subsubsection{Unigram sample complexity analysis} In order to gain quantitative insight into how long the ciphertext needs to be before it can be broken with some desired degree of confidence via unigram analysis and how this minimum length increases with increasing ciphertext alphabet size $|\mathcal{C}|$, we consider the following simpler task for the attacker:  in a binary plaintext alphabet $\mathcal{M} = \{0,1\}$, decide whether a particular ciphertext character $j$ corresponds to the plaintext character $0$ or the plaintext character $1$. Let $n_j$ denote the number of ciphertext characters that equal $j$ in a message of length $n$. For analytical tractability, here we will assume that the plaintext process is stationary and memoryless, i.e., it is a sequence of independent and identically distributed (iid) characters. Then $n_j$ will have a binomial distribution for $n$ trials with success probability equal to $q_0$, if $j$ corresponds to plaintext character $0$, and success probability $q_1$, if $j$ corresponds to plaintext character $1$. The attacker's task of deciding $0$ or $1$ based on $n_j$ and knowledge of $n, q_0, q_1$ is 
a simple Bayesian binary hypothesis testing problem that has been extensively studied in the literature.

Indeed, for sufficiently large $n$, the error probability $P_e(n)$ of the optimum plaintext decoding (\i.e., based on the Maximum A posteriori Probability [MAP] rule) approximates $e^{-n D(r||q_0)}$, where $D(u||v)$ denotes the Kullback-Leibler (KL) divergence~\cite[Section 11.9]{cover2012elements}. 
We have the following result whose proof may be found in Section 11.9 of \cite{cover2012elements}.
\begin{proposition}\cite{cover2012elements} If $\mathcal{M} = \{0,1\}$ and the plaintext process is stationary and memoryless, then for each ciphertext character $j\in \mathcal{C}$, the error probability $P_e(n)$ of the optimum plaintext decoding rule (\ie, the MAP rule) goes to zero exponentially fast with the ciphertext size $n$:
\[
\lim_{n\rightarrow \infty}-\frac{1}{n} \ln P_e(n)
= D(r||q_0) = D(r||q_1)
\]
where 
\[
r(q_0,q_1) := \frac{\ln(1-q_0) - \ln (1-q_1)}{\ln(1-q_0) - \ln (1-q_1) + \ln q_1 - \ln q_0} \in [0,1]
\]
is a probability and $D(u||v) := u\ln(u/v) + (1-u)\ln((1-u)/(1-v))$ denotes the
Kullback-Leibler (KL) divergence between the binary probability
distributions $(u,(1-u))$ and $(v,(1-v))$.\footnote{To be technically
  precise, $D(u||v)$ is finite if $u = 0$ (resp.~1) whenever $v= 0$
  (resp.~1) and is infinity otherwise. Also $0\ln(0)$ is treated as
  zero.}
\end{proposition}

Thus for all $n$ sufficiently large, $P_e(n) \simeq e^{-n D(r||q_0)}$.
In order to achieve a target decoding error probability of $\epsilon$ or less, we require a message of length $n \geq n_{|\mathcal{C}|} = (\ln(1/\epsilon))/D(r(q_0,q_1)||q_0)$ characters. If there was no ciphertext alphabet expansion, i.e., $|\mathcal{C}| = |\mathcal{M}| = 2$, then the minimum number of samples needed to attain a decoding error probability of $\epsilon$ is given by $n_{|\mathcal{M}|} = (\ln(1/\epsilon))/D(r(p_0,p_1)||p_0)$. Therefore, for each $\epsilon$, we need $\frac{n_{|\mathcal{C}|}}{n_{|\mathcal{M}|}}$ times more ciphertext samples compared to the case when there is no 
alphabet expansion. 
\begin{theorem} If $|\mathcal{M}| = 2$ and the plaintext process is iid, then 
the ratio of the length of ciphertext needed to break the cipher (via unigram analysis) with ciphertext alphabet expansion to the length needed without alphabet expansion is given by:
\[
\frac{n_{|\mathcal{C}|}}{n_{|\mathcal{M}|}} = \frac{D(r(p_0,p_1)||p_0)}{D(r(q_0,q_1)||q_0)}.
\]
\end{theorem}

\subsubsection{Example} Consider as a toy example the non-uniform plaintext unigram distribution given by $p_0 = 0.1$ and $p_1 = 1 - p_0 = 0.9$, and the ciphertext alphabet size is $|\mathcal{C}| = 512$ (giving a ciphertext to plaintext size ratio equal to that of UTF-16 Unicode to ASCII). Then taking $|\mathcal{C}_0| = \lceil0.1 \times 512 \rceil = 52$  and $|\mathcal{C}_1| = 512 - |\mathcal{C}_0| = 460$, we get $q_0 = 0.1/52 = 1.9231 \times 10^{-3}$ and $q_1 = 0.9/460 = 1.9565 \times 10^{-3}$ which is more uniform.
Of course, in this particular example if $|\mathcal{C}|$ was a multiple of 10, then the distribution would be exactly uniform, i.e., $q_0 = q_1 = 1/|\mathcal{C}|$ and the cipher will be unbreakable even via multigram analysis (for an iid plaintext process). 
Continuing, we have $r(p_0,p_1) = 0.5, D(r(p_0,p_1)||p_0) = 0.5108$ and $r(q_0,q_1) = 1.9398 \times 10^{-3}, D(r(q_0,q_1)||q_0) = 7.2221 \times 10^{-10}$. This makes $\frac{n_{|\mathcal{C}|}}{n_{|\mathcal{M}|}} = 7.0731 \times 10^6$, i.e., the ciphertext length needed to break a single character (at any confidence level) with a $256$-fold ciphertext alphabet expansion is about $7$ million times that needed to break a single character (to the same confidence level) without ciphertext alphabet expansion. Specifically, for $\epsilon = 0.01$ (99\% confidence), $n_{|\mathcal{M}|} = 9$ and $n_{|\mathcal{X}|} = 63$ million.
We would like to emphasize that these numbers are just for the toy example where the plaintext alphabet has only two characters and the unigram distribution of the two characters is highly non-uniform. These numbers can be expected to be much more larger in practice because typical plaintext alphabet sizes are much larger than $2$ ($95$ for ASCII) and the unigram plaintext distribution is much less skewed.\\

\begin{figure}[t]
\centering
\includegraphics[width=\columnwidth]{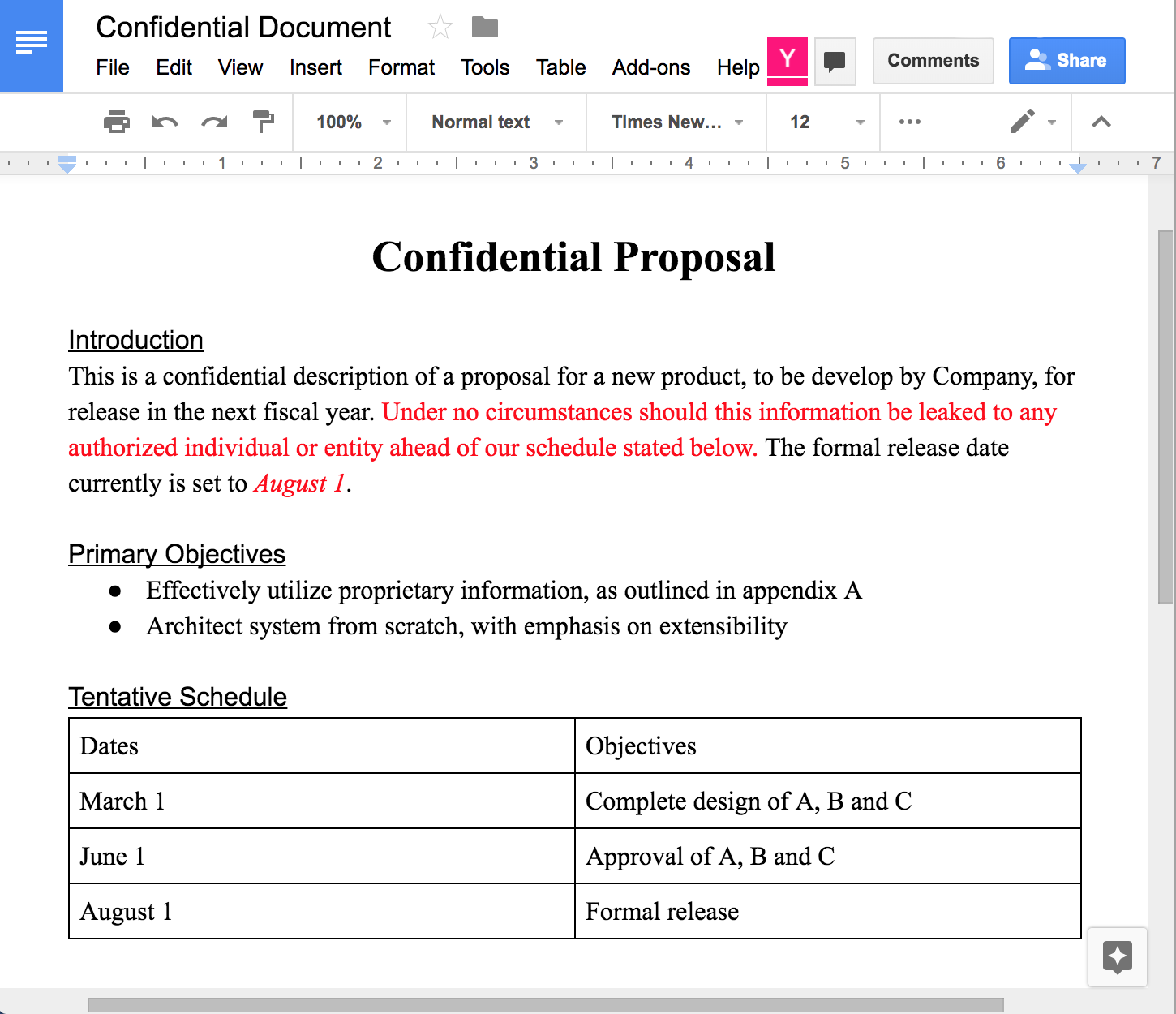}
\caption{A collaborator running our browser extension (with correct password) edits a Google Doc. The app functions normally with its collaborative features. However, data specifying the document's text contents is intercepted and modified in transit to/from Google's servers according to our encryption. Only users utilizing the correct password can view the original contents.
}
\label{demo:plain}
\end{figure}

\subsection{Plaintext Attacks}
\label{subsec:plaintext}
A known or chosen plaintext (with a corresponding ciphertext) significantly reduces the complexity of breaking a substitution cipher by providing some of the plaintext-ciphertext substitutions that form the encryption key; language and context may be used to infer the remaining substitutions.  Utilizing a polyalphabetic cipher, as described in this work, improves the resilience of the cipher, since less information is revealed with each substitution.  In other words, if the letter character \texttt{a} is mapped uniformly at random to one of ten ciphertext characters, then revealing one of these plaintext-ciphertext connections only reveals one tenth of the occurrences of \texttt{a}.  This mapping can be modified in some coarse manner, say based on the month in which the text is produced or the name of the original author of the work, in order to limit the usability of known plaintexts.

The cipher can be made even more resilience by encrypting one plaintext character with more than one ciphertext character, and, indeed, this does not break our collaborative encryption model or our prototype implementation, although it is possible that location-sensitive processing could suffer. 
Though character-level encryption is inherently weaker than block- or stream-based encryption, we stress that the proposed approach provides a measurable increase in privacy, where currently none exists, without requiring server-side or browser rewriting, and that the encryption can be strengthened further at the expense of efficiency.

\section{Prototype}
\label{sec:demo}
We demonstrate our proposed encryption framework through the Google Docs platform on
the Google Chrome. 
Google Docs is a collaborative document editing service provided by Google.
Two or more users can edit a document's state (\ie text, formatting, and figures)
together in real-time using only a modern web browser. Data resides on Google's servers, and any collaborative edits pass through Google's infrastructure before being forwarded to other collaborators.

\begin{figure}[t]
\includegraphics[width=\columnwidth]{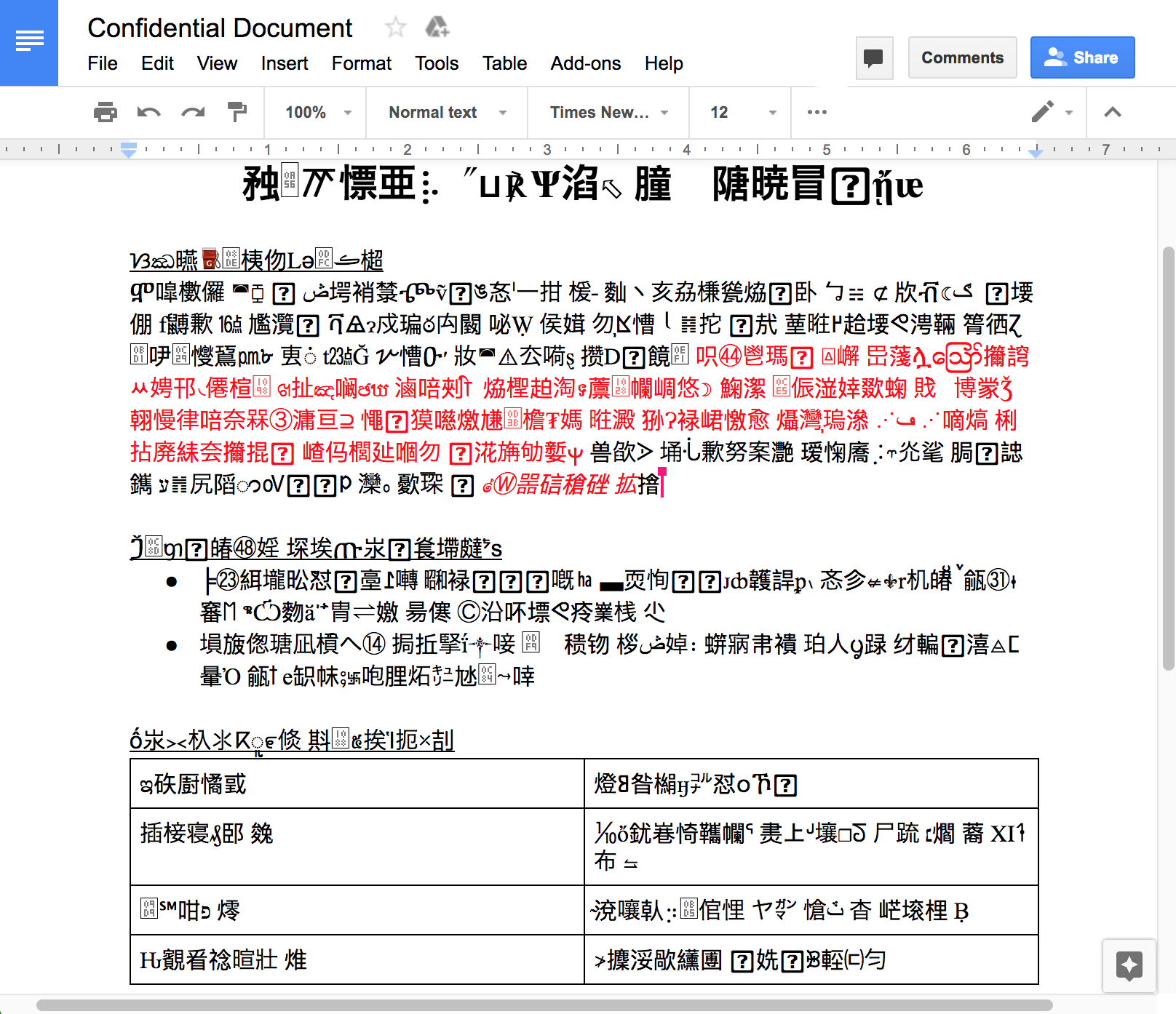}
\caption{An undesired third party (or the provider) views the Google Docs document as a guest without a correct private key. Since the document has been configured to allow public viewing, Google Docs permits access. However, in the absence of the extension and a valid password, document text is indecipherable to the third party.
}
\label{demo:cipher}
\end{figure}

\subsection{Mechanism}
For Google Docs, document data is structured as a series of event messages, each of which has an associated opcode and a set of fields specific to that opcode. The client-side app parses these messages, which specify document contents, layout, and formatting, to render a document for the user.

With our architecture we were able to intercept both sources of information using the techniques in Section~\ref{sec:browser-interface}. First 
our implementation involves hooking the \texttt{XMLHttpRequest} prototype for all frames originating from Google's relevant servers, providing access to the incoming and outgoing XHR data streams. By examining the effects on document state of messages with particular opcodes, we found that events with opcode \texttt{is} specify one or more character insertions. By filtering for \texttt{is} events, we captured collaborative edits of a document's text contents.

We then exploited the \texttt{document\textunderscore start} feature to gain control of the \texttt{DOCS\textunderscore modelChunk} variable at app load-time and inspected the series of event messages it contained. We found that these messages captured the cumulative document state of all previous collaborative edits. The messages followed the same format as those of the XHR stream, but state changes were combined across messages where possible. Filtering for \texttt{is} events again provided enough specificity to capture all document state information concerning text contents,
because messages specifying formatting state such as text size, color, and layout have an opcode other than \texttt{is}.

\subsection{Performance Evaluation}
We focused our evaluation of the project demonstration on users' experience in both qualitative and quantitative way. 

Qualitatively, our prototype extension does not modify formatting (bold type, font size, line spacing, etc.) as well as the Google Docs GUI. This enables users to edit and collaborate as if the encryption layer does not exist. The only discrepancy is spelling corrections (which are handled server-side) are disabled as servers only store ciphertext. If the user mistypes a word, the word would not have an red underline or any suggestion for correction. 

On the quantitative side, we measured the delays our extension would cause due to encryption, recording the time between the start and the end of overridden XMLHttpRequest call.  The resulting delay is linearly proportional to the number of characters being encrypted, fitting the equation $t = 0.0017n + 6.0876$, 
where $n$ is the number of character edits, and time $t$ is measured in milliseconds and measures both outgoing and incoming data stream. 
Since large numbers of character edits happen only during the process of copy-and-paste and loading of the page, most users will experience an average of $6ms$ delay during their active editing. Because the delay is not significant, we conclude that such delay does not affect the overall users' experience.

\section{Conclusions}
\label{sec:conclusions}
We have presented a client side encryption system for real time collaborative editing web app. The system consists of an encryption interface as well as a novel variant of the polyalphabetic substitution cipher, designed to seamlessly encrypt and decrypt data without interfering with app functionality or the users' experience within the web app.  In this way, the users' data privacy is preserved both during transmission and at rest with the provider.

We have implemented a prototype of our system for the Google Docs collaborative word processing web application within the framework of the Google Chrome web browser.  We believe that our choice of prototyping framework serves as a reasonable template for other major browsers and web apps, and that our design can be straightforwardly extended to any real-time collaborative editor which uses the standard \texttt{XHR} interface for client-server communication, including such well-used products as the Google productivity suite (Docs, Slides, Sheets), Conceptboard, MeetingWords, Collabedit, Codepen
\cite{conceptboard,meetingwords,collabedit}.

For potential extensions to this work, minor modifications need to be made in order for the extension to work with different browsers. Moreover, an automated process of recognizing formatting of event messages can be developed in order to avoid manual analysis for different web apps. Finally, with respect to encryption, a motivated provider could identify and blacklist features of our approach, such as the utilization of a broad range of the Unicode spectrum.

\section*{Acknowledgments}
The authors acknowledge John Navon and John Moore for work on an earlier versions of this work, and Manuel Egele for early proofreading and feedback.  This work was also supported, in part, by the National Science Foundation under Grant No. CCF-1563753.

\bibliographystyle{abbrv}

\balance  

\end{document}

%% file: macros.tex
\newcommand{\ie}{\emph{i.e., }}
\newcommand{\eg}{\emph{e.g., }}
\newcommand{\itembf}[1]{\item \textbf{#1}}
\newcommand{\itemit}[1]{\item \emph{#1}}
\newcommand*\q{\mathchar`'}

\newcommand*\M{\ensuremath{\mathcal M}}
\newcommand\C{\ensuremath{\mathcal C}}

\definecolor{lightgray}{rgb}{.9,.9,.9}
\definecolor{darkgray}{rgb}{.4,.4,.4}
\definecolor{purple}{rgb}{0.65, 0.12, 0.82}

\setlength\itemsep{1em}
\setlength\partopsep{1em}